# Multifractal analysis of Mt. St. Helens seismicity as a tool for identifying eruptive activity

Filippo Caruso [1]

Scuola Superiore di Catania, Università di Catania, Via S. Nullo, 5/i, I-95123 Catania, Italy

Sergio Vinciguerra

HP-HT Laboratory of Experimental Volcanology and Geophysics, Department of Seismology and Tectonophysics, INGV, I-00143 Rome, Italy

Vito Latora, Andrea Rapisarda

Dipartimento di Fisica e Astronomia, Università di Catania, Via S. Sofia 64, I-95123 Catania, Italy

Stephen Malone

Department of Earth and Space Sciences, University of Washington, Seattle, WA 98195, USA

**Abstract**

We present a multifractal analysis of Mount St. Helens seismic activity during 1980-2002. The seismic time distribution is studied in relation to the eruptive activity, mainly marked by the 1980 major explosive eruptions and by the 1980-1986 dome building eruptions. The spectrum of the generalized fractal dimensions, i.e. $D_q$ vs q , extracted from the data , allows us to identify two main earthquake time-distribution patterns. The first one exhibits a multifractal clustering correlated to the intense seismic swarms of the dome building activity. The second one is characterized by an almost constant value of $D_q \approx 1$, as for a random uniform distribution. The time evolution of $D_q$ (for q=0,2), calculated on a fixed number of events window and at different depths, shows that the brittle mechanical response of the shallow layers to rapid magma intrusions, during the eruptive periods, is revealed by sharp changes, acting at a short time scale (order of days), and by the lowest values of $D_q$ ($\approx 0.3$). Conversely, for deeper earthquakes, characterized by intense seismic swarms, $D_q$ do not show obvious changes during the whole analyzed period, suggesting that the earthquakes, related to the deep magma supply system, are characterized by a minor degree of clustering, which is independent of the eruptive activity.

1) Present address: Scuola Normale Superiore, Piazza dei Cavalieri, 7
   I-56126 Pisa, Italy ; Email : filippo.caruso@sns.it

# 1. Introduction

Earthquakes space-time distributions exhibit fractal properties[1] that can be a consequence of a self-organized critical state of the earth crust, analogous for instance to the state of a sandpile that spontaneously evolves to a critical angle of repose in response to the steady supply of new grains at the summit[2,3]. The self-similar nature of seismicity in space and time is then an indication of the hierarchical structure of the clustering, intimately linked to the stress in the earth's crust. Consequently, the numerical extraction of the fractal dimensions of real seismic data is an important method to quantify the distribution of seismic events and with that the properties of randomness and clusterization in a given geographical region. In particular, considering the evolutionary aspect, the study of the temporal variations of the spatial fractal dimension can show how the global scaling relations, namely the space scale invariance, change in time[4,5,6]. The time evolution of the fractal dimension provides more information than simpler analysis, as for example the variation of the number of events inside a fixed time window, being able to distinguish and characterize seismic patterns acting at different time scales. In particular, in volcanic environments time variations of time-space fractal dimensions and b-value have been interpreted in terms of rapid changes of the physical state of the volcano edifice acting at mid (order of months) and short-term (order of days) scales[7,8,9]. Nevertheless, recent studies suggest that multifractal laws, rather than simple monofractal ones, are necessary to explain the scale-related complexity observed in earthquake distributions and their time evolution[6,10,11].

In the present work we study the multifractal properties of the Mount St. Helens volcano seismicity distribution during the period 1980-2002. We focus on the time distribution of seismic events, which is the most important marker of the changes of the physical state of the volcano. We therefore present an analysis aiming at defining how the properties of seismic time clustering evolve in time with respect to: 1) eruptive and different phases of dome building activity; 2) relative quiescence phases of the volcano; 3) different depths of crustal volumes, whose mechanical behavior is controlled by the interplay of regional tectonic and magma stresses. We extract the generalized fractal dimensions from the time sequences and discuss how the multifractal formalism is a well suited tool for volcanic areas in which multiple processes act. We also discuss how in this way one can extract unique information by analyzing only earthquakes time occurrence, which may constitute a major support for monitoring purposes.

## 2. Mount St. Helens volcanism and seismicity

Mount St. Helens, Washington, is the most active volcano in the Cascade Range. The volcano awakened in mid-March[12] of 1980 with an intense, swarm of earthquakes preceding violent phreatic eruptions and rapid northward deformation of the north flank of its volcano beginning no later than early April. By May 17 more than 10,000 earthquakes, mostly shallow (with a depth $h<3$ km), had occurred, including a magnitude 4.2 earthquake on March 20. Meanwhile, the north flank of the volcano had grown outward more than 140 meters. On the morning of May 18, a magnitude-5.1 earthquake triggered a cataclysmic lateral blast destroying everything up to 30 km north of the volcano, followed by pyroclastic flows and volcanic mudflows (lahars) lasting several hours. From May 25 to October 16 five smaller explosive episodes occurred, producing eruption columns and pyroclastic flows. After the May 18 plinian eruption, earthquakes were much smaller than before it. Following each explosive eruption a deep ($h>4$ km) suite of earthquakes occurred.

The character of the seismicity and eruption patterns changed dramatically after the explosive eruption of October, 1980. Primarily episodic dome-building eruptions took place from October 1980 to 1986, each preceded by a precursory swarm of earthquakes ( see Fig. 1) and minor deformation within the crater. It is remarkable that, during this period, 'deep' earthquakes cluster hypocenters become 'deeper' between 6 and 11 km, while the bulk of 'shallow' seismicity affects depths between 0 and 4 km.

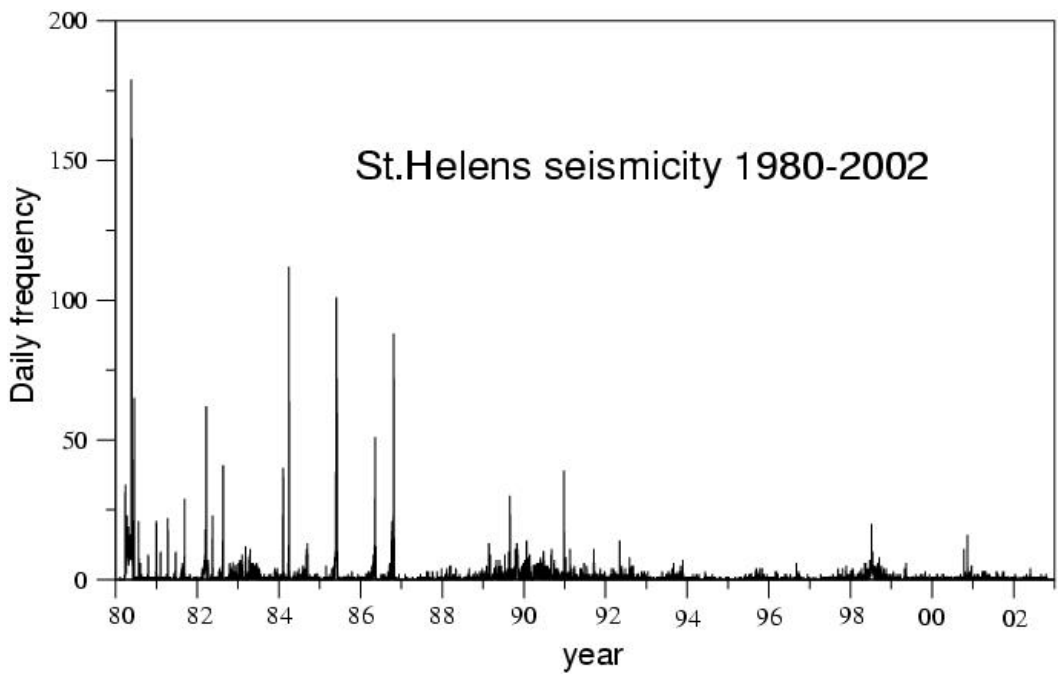

*Fig.1 - Daily frequency of Mt. St. Helens seismicity in the period 1980-2002.*

After the end of dome-building eruptions in October, 1986 seismicity at St. Helens decreased considerably through 1989. Shallow (h<3 km) seismicity has been mostly absent since then. Starting in 1989 persistent patterns of deeper earthquakes cluster in the 4-10 km zone were observed. Between 1989 and 1991 a number of swarms of deeper seismic activity accompanied small steam explosions from the dome. Overall seismicity in the shallow, 2-4 km zone, is of low levels and not concentrated in time. Since 1991 no explosions have occurred and seismicity in the 0-2 km zone is almost absent. The 2-4 km deep zone has persistent low level of earthquakes. Episodes of significantly increased seismicity occurred in both 1995 and 1998, primarily in the deeper zone of 4 to 8 km depth. The increased deep seismicity around 1989-1991, in 1995 and in 1998 has been interpreted, based on focal mechanisms, as being caused by stress release on local faults and fractures due to magma refilling the crustal magma chamber[13].

## 3. Multifractal dimensions

The fractal dimension is the basic concept to describe structures having a scaling symmetry. Scaling symmetry means self-similarity of the considered object on varying scales of magnification. Historically the first definition of generalized fractal dimension, providing a measure for filling space which allows for the possibility of non-integral dimensions, was introduced by Hausdorff in late 1919. The capacity-dimension $D_C$ of a given geometric object (set), also called the *box-counting* dimension[15,16], is defined in the following way. First one has to perform a partition of the space occupied by the object into N equally sized d-dimensional cubes with edge size $\varepsilon$, and then one counts the minimum number of cubes $N(\varepsilon)$ required to cover the set. $D_C$ is a measure of how $N(\varepsilon)$ scales with $\varepsilon$:

$$D_C = -\lim_{\varepsilon \to 0} \frac{\ln N(\varepsilon)}{\ln \varepsilon} \quad . \quad (1)$$

Such a measure, only based on the number of occupied boxes, does not take into account the possibility of having fractal regions with different density. In general, to characterize completely a distribution, or to distinguish between two distributions quantitatively, we need to compare the different moments of the distribution. In a similar way the geometry of the fractal object under study is better characterized by means of an infinite set of generalized fractal dimensions.

The multifractal dimensions of q-th order[14,15,16] are defined as:

$$D_q = \lim_{\varepsilon \to 0} \frac{1}{q-1} \frac{\ln\left(\sum_{i=1}^{N} p^q_i(\varepsilon)\right)}{\ln \varepsilon} \quad q = -\infty, \ldots, +\infty \quad . \quad (2)$$

As before, one has to divide the space, the support of measure μ, into N d-dimensional cubes of size ε denoted by $\Lambda_i$ (i=1,..,N). We indicate with $p_i(\varepsilon) = \int_{\Lambda_i} d\mu$ the integrated measure on the i-th cube, i.e. the mass of $\Lambda_i$. In practice, if the fractal is made by a set of M points (each point corresponds to the occurrence of an earthquake in the case under study), $p_i(\varepsilon)$ is given by the probability of finding a point in the i-th box $p_i(\varepsilon) = \frac{M_i(\varepsilon)}{M}$ ( $M_i$ being the number of points in $\Lambda_i$). It is important to realize that the generalized fractal dimension involves the probability raised to the q-th power $p^q_i(\varepsilon)$. Thus the multifractal dimension $D_q$ weights in a different manner the various density regions. In particular the limiting dimension $D_{-\infty}$ and $D_{+\infty}$ are related to the regions of the set in which the measure is most dilute and most dense respectively. Using this definition, one finds as a special case the previously defined box-counting fractal dimension for q=0, i.e. $D_0=D_C$. On the other hand, for q=1, one gets the so-called *information* dimension $D_I$ :

$$D_1 = D_I = \lim_{\varepsilon \to 0} \frac{\left(\sum_{i=1}^{N} p_i(\varepsilon) \ln p_i(\varepsilon)\right)}{\ln \varepsilon} \quad . \quad (3)$$

While, for q=2, we get the *correlation* dimension $D_G$[14] :

$$D_G = \lim_{\varepsilon \to 0} \frac{\ln C(\varepsilon)}{\ln \varepsilon}, \quad (4)$$

$$C(\varepsilon) = \frac{1}{M \cdot (M-1)} \sum_{\substack{i,j=1 \\ i \neq j}}^{M} \theta\left(\varepsilon - |\vec{x}(i) - \vec{x}(j)|\right), \quad (5)$$

where by $\vec{x}(i)$ $(i=1,...,M)$ we indicate the positions of the M points of the object under study, and θ is the Heaviside function.

In general we have $D_q \geq D_{q'}$ for $q < q'$. Only in the very particular case of an object with equal probabilities for all the cells, i.e. for a *monofractal*, we have $D_q = D_0$ for all q.

In order to calculate the complete spectrum, one can also write eq. (2) by means of the generalized correlation sum $C_q$ [15,16] as:

$$D_q = \lim_{\varepsilon \to 0} \frac{1}{q-1} \frac{\ln C_q(\varepsilon)}{\ln \varepsilon}, \quad (6)$$

with

$$C_q(\varepsilon) = \frac{1}{M} \sum_{i=1}^{M} \left[ \frac{1}{M-1} \sum_{\substack{j=1,\\j \neq i}}^{M} \theta\left(\varepsilon - |\vec{x}(i) - \vec{x}(j)|\right) \right]^{q-1}. \quad (7)$$

The main goal of this paper is to compute the multifractal dimensions $D_q$ of the time distribution of earthquakes. For our purpose the series $\vec{x}(i)$ $(i=1,...,M)$ is given by t(i) ( i=1,…,M), i.e. by the time occurrence of the M earthquakes.

## 4. Data analysis and discussion

Seismic data are from the master catalog of the Pacific Northwest Seismograph Network, (http://www.pnsn.org/CATALOG_SEARCH/cat.search.html) relatively to St. Helens coordinates (46.17 < Latitude < 46.23; 122.14 < Longitude < 122.23) (Fig. 1 reports the daily distribution). A catalogue of around 9000 events has been downloaded.

Two main phases can be distinguished in terms of volcanic activity: the 1980-1986 eruptive period, which includes the cataclysmic May 18, 1980 and 1981-1986 dome building period, and the 1987-2002 period with no eruptive activity, but only a few very small steam explosions in the winter 1989-1990, 1991 characterized by the absence of juvenile material.

**4.1 Multifractal spectra for eruptive and non eruptive periods**

In Fig. 2(a) we report the spectrum of the generalized fractal dimensions $D_q$ obtained for the eruptive and for the non eruptive period. In the case of the eruptive period (1980-1986), the spectrum clearly indicates that the time distribution of earthquakes has a multifractal structure with generalized fractal dimensions ranging from $D_q \sim 0.36$ for $q \to \infty$, to $D_q \sim 1.0$ for $q \to -\infty$. This fact means a presence of strong clustering with regions of different density in the time distribution.

Conversely, when $D_q$ are computed for the non eruptive period (1987-2002), we measure a spectrum of $0.7 \leq D_q \leq 1$, indicating a smaller range and higher values. This fact suggests that seismicity is distributed more uniformly during this period and that the time series is closer to a homogeneous random sequence with respect to the eruptive case.

It is important to observe that, by using the spectrum of multifractal dimensions, we can better discriminate between eruptive and non eruptive periods than by means of $D_0$ only. In fact, for example considering only q=0, we have $D_0$ =0.97 and $D_0$=0.87 for the non eruptive and eruptive periods respectively. However, one could also have objects with the same $D_0$ but different spectra. So, in general, one should always prefer a multifractal analysis to a simple calculation of the box-counting dimension in order to characterize in a unique and quantitative way a multifractal object[15,16].

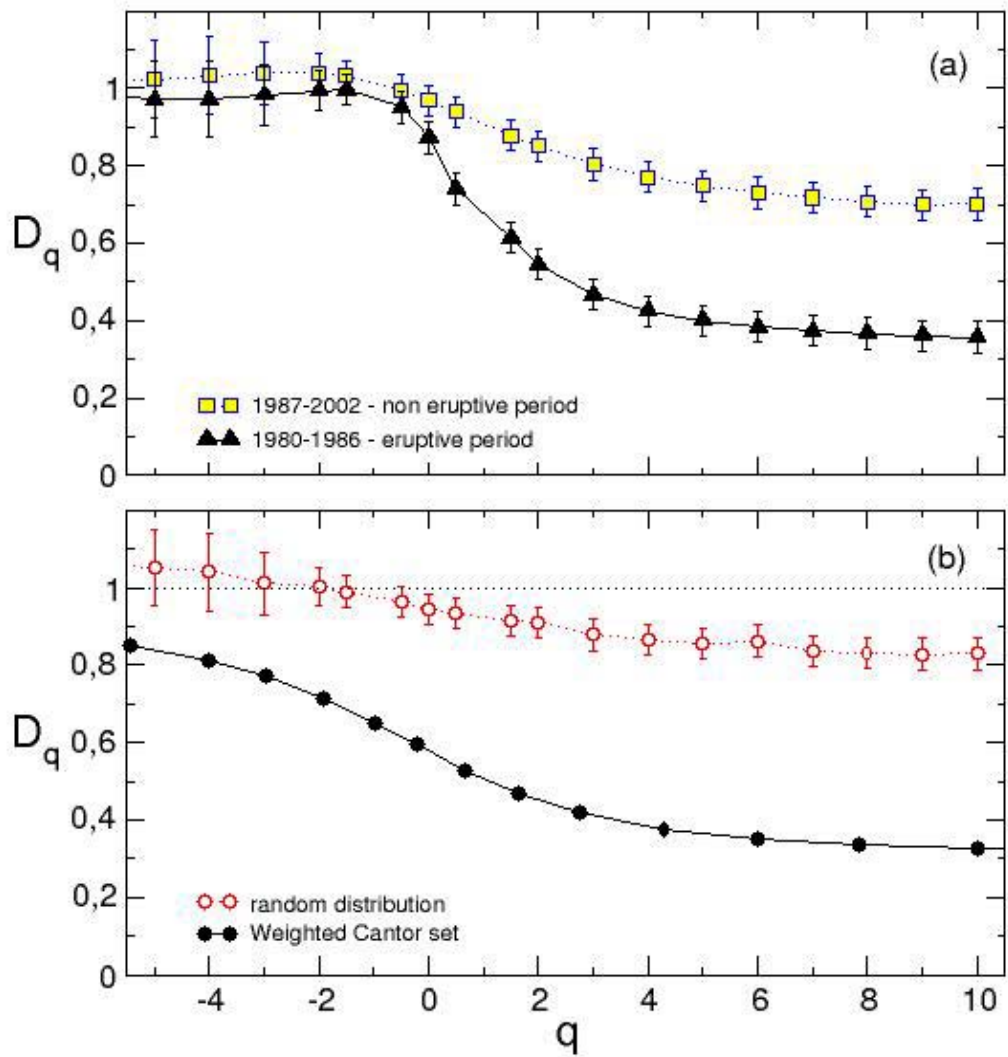

*Fig.2 - (a) The spectrum of generalized fractal dimensions is reported for the two periods of the database: the eruptive one 1980-1986 (full triangles) and the non eruptive one 1987-2002 (squares) . (b)We plot the spectra of generalized fractal dimensions for an artificial multifractal set, the weighted Cantor set , with $p_1$=0.7 and $p_2$= 0.3 (see Ref.15 p.400 or Ref.16 p.340) (full circles), and for a uniform random distribution of 1700 events in a period of 10 years (empty circles) as for the real data in the non eruptive period.*

## 4.2 Multifractal analysis of artificial sequences

In order to better understand the results of our analysis, we have computed the spectra of generalized fractal dimensions for two artificially generated distributions. The first artificial sequence was obtained by homogeneously randomly distributing the same number of earthquakes present in the non eruptive period. The second one refers to a well known multifractal object, the *weighted Cantor set*, a typical multifractal set obtained by using two different probabilities $p_1$=0.7 and $p_2$=0.3 for constructing a generalized fractal Cantor set with different population densities (see for example Ref. 15 p.400 or Ref. 16 p.340 for more details). Figure 2(b) shows that the results are perfectly compatible with the behavior obtained for the non eruptive and eruptive period respectively.

In particular the uniform random distribution does not give exactly $D_q$=1 for all q, as one would expect, because of the finiteness of the sample, however it is very similar to the shape observed in the non eruptive case. This fact reinforces our claim that the seismic activity is randomly distributed in time when no eruptive activity is present, while the seismic activity is clustered in time (as in a multifractal set) when an eruptive process is active.

## 4.3 Time evolution analysis of the generalized fractal dimensions

Time evolution of the multifractal spectrum was investigated in order to identify and/or distinguish seismic patterns acting at different time scales. After a few tests, we adopted a moving window of 250 events with an overlap of 50 events. An example of typical plots for extracting the two dimensions $D_0$ and $D_2$ is reported in Fig.3. The figure shows that the number of events considered for each window is sufficient for a good linear fit and a reliable estimate of the fractal dimensions. We report in Fig. 4(b) the time evolution of the dimensions $D_2$ and $D_0$. Both of them show sharp changes acting at short time scales (order of months or days) controlled by the seismicity time distribution of the 1980-1986 eruptive period. We show the magma volume erupted (in million cubic yards) in Fig. 4(a).

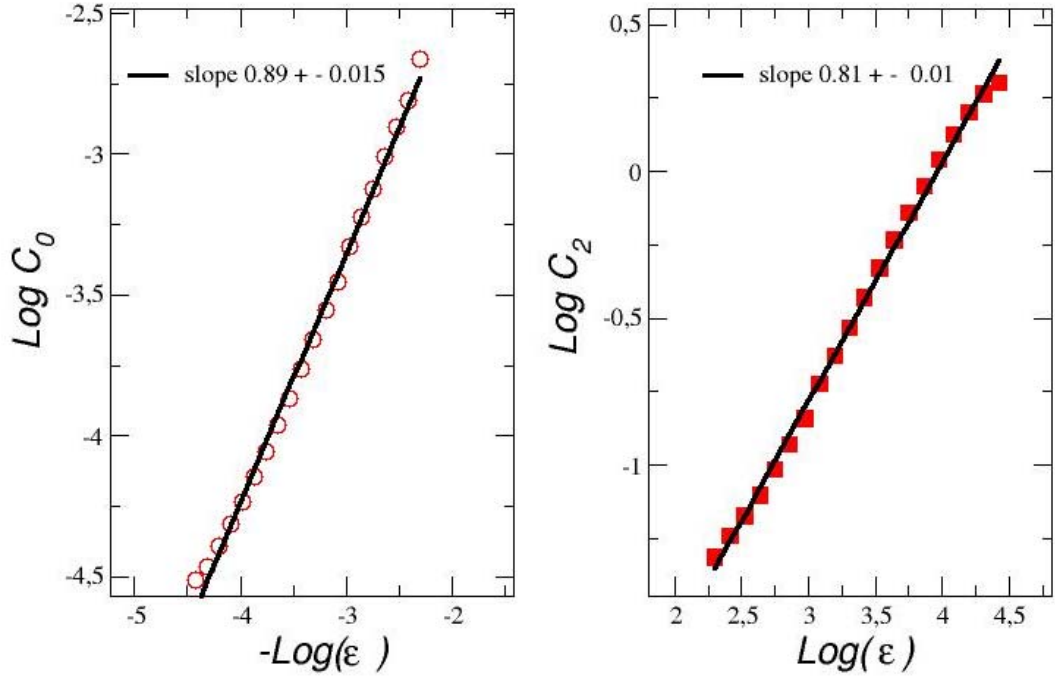

*Fig.3 – Examples of the typical plots used to extract the generalized fractal dimensions reported in Fig.4 with 250 events per window. This case refers to the 10$^{th}$ window. The full line is a linear fit of the data. The value of the slopes are also reported with the corresponding error estimate.*

This behavior, induced mainly by intense seismic swarms related to the dome-building activity, determines a large variability in the multifractal spectrum, with the lowest values of $D_2 \approx 0.3$-$0.4$, marking high values of fractal clustering. $D_q$ variability can be related to the brittle response of the medium by means of intense seismic swarms during the dome building activity and the rise of fresh magma feeding in the shallower portion of the crust, eventually causing large volume eruptions[17]. $D_q$ time evolution is much more stable ($0.6 \leq D_{0,2} \leq 0.9$) during 1987-2002, where only a sharp decrease occurred in 1991 ($D_2 \sim 0.5$), probably related to the renewal of small explosions from the dome and related seismic swarms. The major implication is that this period is mostly characterized by the 'background' seismic activity, i.e. the regional or tectonic seismic activity, which shows a much more stable behavior of $D_q$ and its higher values, tending towards a random uniform distribution as shown in Fig. 2.

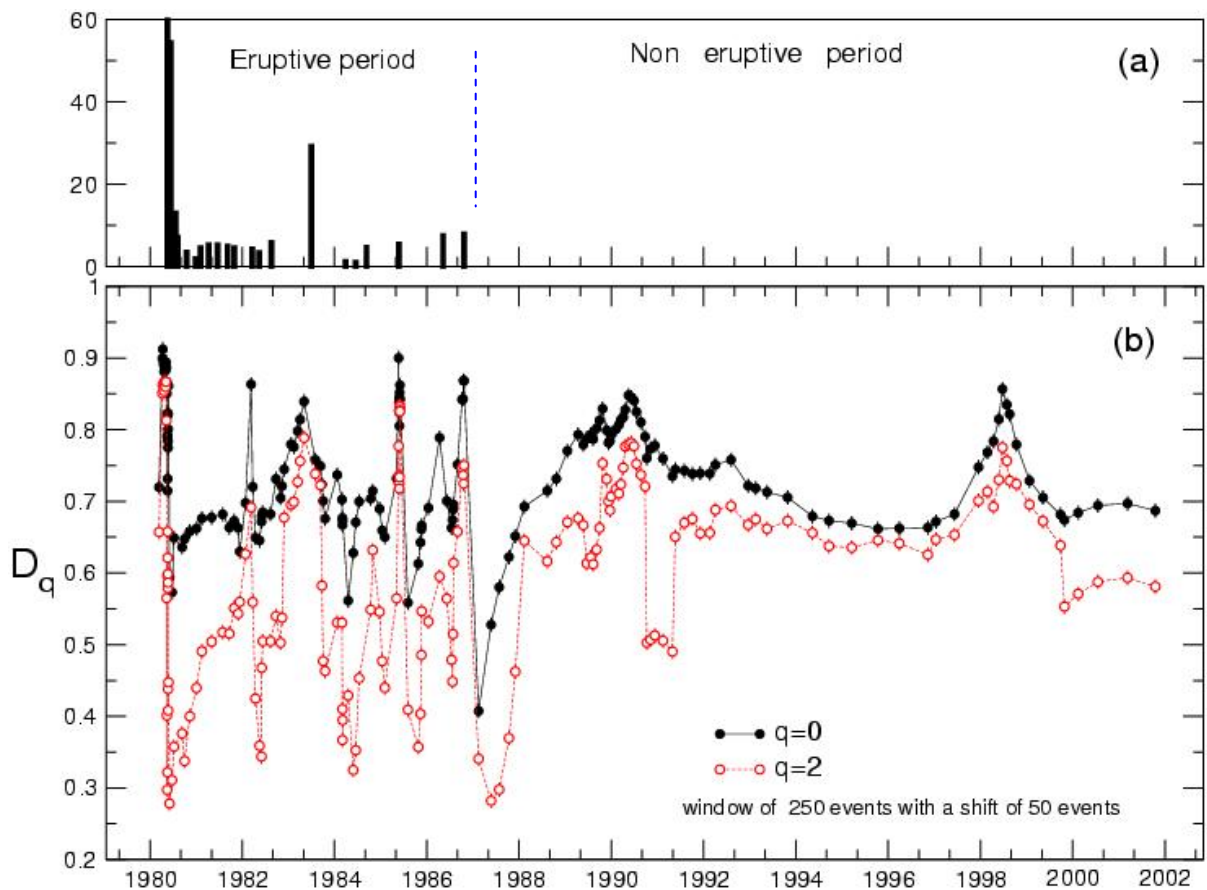

*Fig.4. a) We report the eruptions occurred in the studied period. Bars indicate the magma volume erupted in million cubic yards.*
*b) Here we plot the generalized fractal dimension $D_0$ and $D_2$ calculated for a moving window in the time sequence consisting of 250 events with a shift of 50 events. In the eruptive period a close correlation between rapid changes in the value of the two dimensions with the major eruption is clearly visible. The time evolution of the fractal dimensions becomes smoother after the dome-building activity period.*

## 4.4 Depth analysis of seismicity

In order to take into account the different mechanical behaviour of the crust below St. Helens, $D_q$ has been also calculated according to different depths. On the basis of seismo-tectonic and volcanological consideration[12], two main crustal volumes below St. Helens are considered. The first, located between 0 and 4 km of depth, is made of low-strength volcanic rocks, which failed either due to the dome building activity or to magma rise from the depth. The second, located between 4 and 10 km of depth, is mostly related to regional or tectonic stresses interplayed with magma overpressures within the mid-crust. Magma may move into zone 2 from a deep source slowly and perhaps continuously[12] with pulses or increased flow from time to time[13].

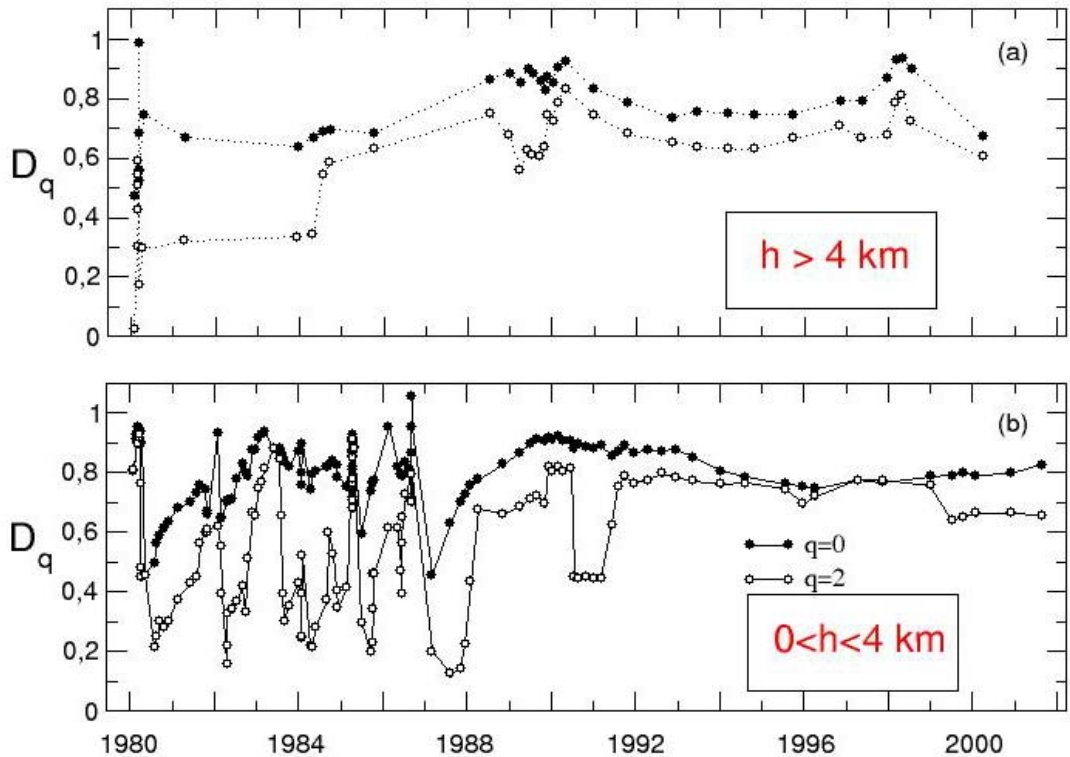

*Fig.5 – Here we plot the generalized fractal dimension $D_0$ and $D_2$ calculated for a moving window in the time sequence as in the previous figure, considering only seismic events registered in the h>4 km depth zone (a) and earthquakes in the 0-4 km depth zone (b). See text.*

The behavior of $D_{0,2}$ as a function of time and for these two regions of depth is plotted in Fig.5. Here one can see that the $D_q$ calculated for the 0<h<4 km zone shows the same trend reported in Fig. 4, evidencing high variability during 1980-1986, where the lowest values of $D_q$ ($\approx 0.3$) are found and a more stable trend during 1987-2002 ($0.6 \leq D_q \leq 0.9$). This is a further confirmation that large fluctuations of $D_q$ are induced by the brittle mechanical response of the shallow layers to rapid magma intrusions marking eruption onset. When $D_q$ are calculated for deeper earthquake foci (h>4 km), a limited number of earthquakes remains available. However, despite the data paucity, a reduced change of $D_q$ over time, during and after the eruptive period, is observed. This implies that the earthquakes, related to the deep magma supply system, show a significantly different fractal clustering and are characterized by an almost random structure, which is almost independent from eruptive activity.

## 5. Conclusions

We have shown that the changes of the volcanic activity observed at Mt. St. Helens can be characterized quantitatively by considering the spectrum of generalized fractal dimensions $D_q$. More precisely, during the eruptive period 1980-1986 we have observed a characteristic multifractal clustering with $D_q$ spanning a range [1.,0.36], while during the non eruptive period 1987-2002, we have obtained a range of variability in the interval [1,0.7], a result which indicates that the time series is much less clustered.

A finer analysis performed by calculating the multifractal dimension $D_q$ in time moving windows has revealed a clear correlation between a gradient in $D_0$ and $D_2$ and the major eruptions occurrence. On the other hand, during the following non eruptive period an almost constant behaviour with $D_0 \approx D_2 \approx 0.8 \pm 0.2$ is found. Such values are consistent with a uniform random signal.

Time evolution calculated with respect to depth reveals that $D_q$ variability is peculiar of shallow crust volumes (0<h<4 km), probably induced by the mechanical response to magma intrusions, while the deeper seismicity is characterized by an almost random structure, which is independent from the eruptive activity.

The method we have presented allows to extract useful information considering only time occurrence and can be a very powerful tool for monitoring issues, especially for the vast majority of active volcanoes which are poorly monitored.

**References**

1. D.L. Turcotte, *Fractals and Chaos in Geology and Geophysics*, Cambridge University Press (1992).

2. P. Bak, C. Tang, K. Wiesenfeld, *Phys. Rev. Lett.,* **59**, 381-384 (1987); H. J. Jensen, *Self-Organized Criticality*, (Cambridge University Press, Cambridge, 1998).

3. D. L. Turcotte, *Phys. Earth Planet. Int*. 111, 275-293 (1999); F. Caruso, A. Pluchino, V. Latora, A. Rapisarda, B. Tadic, *Eur. Phys. Journ. B* **50**, 243-247 (2006).

4. P. Tosi, V. De Rubeis, V. Loreto, L. Pietronero, *Ann. of Geophys*. **47,** 1-6 (2004) ; V. De Rubeis, P. Dimitriu, E. Papadimitriou and P. Tosi, *Geophys. Res. Lett*. **20**, 1911-1914 (1993).

5. V. De Rubeis, P. Tosi, S. Vinciguerra, *Geoph. Res. Lett.* **24**, 2331-2334 (1997).

6. D. Legrand, A. Cisternas, L. Dorbath, *Geophys. Res. Lett.* **23**, 933-936 (1996).

7. S. Gresta and G. Patanè, *Pure Appl. Geophys*. **121**, 287-295 (1983).

8. V. Latora, A. Rapisarda, S. Vinciguerra, *Phys. Earth and Plan. Int.* **109**, 115-127

(1998).

9. S. Vinciguerra, *Geophys. Res. Lett.* , **26**, 24, 3685-3688 (1999).

10. P. P. Dimitriu, E. M. Scordilis, V. G. Karacostas, *Natural Hazards* **21**, 277-295 (2000).

11. L. Telesca, V. Lapenna, F. Vallianatos, *Phys. Earth Planet. Int*. **131**, 63-79 (2002).

12. S. Malone, *Geoscience Canada* **17**, 3, 146-150 (1990).

13. C. Musumeci, S. Gresta, SD. Malone, *J. Geophys. Research-Solid Earth* **107**, 16 (2002).

14. P. Grassberger and I. Procaccia, *Physica D* **9**, 189 (1983).

15. R. C. Hilborn, *Chaos and Nonlinear dynamics*, 2nd ed. (Oxford University Press, Oxford, 2000).

16. J. C. Sprott, *Chaos and Time-Series Analysis*, (Oxford University Press, Oxford, 2003).

17. S. R. Brantley, B. Myers, Mount St. Helens: from the 1980 eruption to 2000, USGS fact sheet 0036-00, http://geopubs.wr.usgs.gov/fact-sheet/fs036-00/fs036-00.pdf (2000).